# Anti-fatigue-fracture hydrogels

Shaoting Lin,[1*] Xinyue Liu,[1*] Ji Liu,[1*] Hyunwoo Yuk,[1] Hyun-Chae Loh,[2] German A. Parada,[1,3] Charles Settens,[4] Jake Song,[5] Admir Masic,[2] Gareth H. McKinley,[1] Xuanhe Zhao[1,2]†

[1]Department of Mechanical Engineering, Massachusetts Institute of Technology, Cambridge, MA 02139, USA

[2]Department of Civil and Environmental Engineering, Massachusetts Institute of Technology, Cambridge, MA 02139, USA

[3]Department of Chemical Engineering, Massachusetts Institute of Technology, Cambridge, MA 02139, USA;

[4]Center for Materials Science and Engineering, Massachusetts Institute of Technology, Cambridge, MA 02139, USA;

[5]Department of Materials Science and Engineering, Massachusetts Institute of Technology, Cambridge, MA 02139, USA;

Shaoting Lin, Xinyue Liu and Ji Liu contributed equally to this work.

†To whom correspondence should be addressed. Email: zhaox@mit.edu

**One sentence summary:** Designing nano-crystalline domains gives extremely anti-fatigue-fracture hydrogels for artificial cartilages and soft robots.

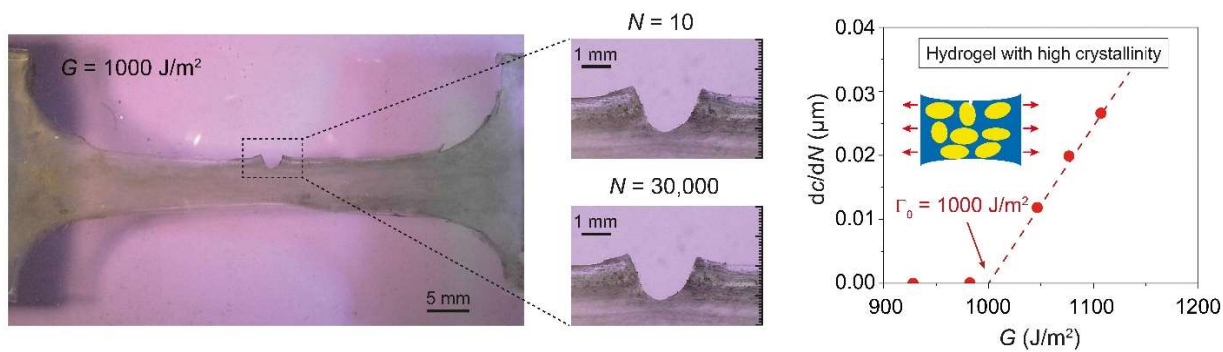

  

**Abstract**

The emerging applications of hydrogels in devices and machines require these soft materials to maintain robustness under cyclic mechanical loads. Whereas hydrogels have been made tough to resist fracture under a single cycle of mechanical load, these toughened gels still suffer from fatigue fracture under multiple cycles of loads. The reported fatigue threshold (i.e., the minimal fracture energy at which crack propagation occurs under cyclic loads) for synthetic hydrogels is on the order of 1–100 J/m$^2$, which is primarily associated with the energy required to fracture a single layer of polymer chains per unit area. Here, we demonstrate that the controlled introduction of crystallinity in hydrogels can significantly enhance their fatigue thresholds, since the process of fracturing crystalline domains for fatigue-crack propagation requires much higher energy than fracturing a single layer of polymer chains. The fatigue threshold of polyvinyl alcohol (PVA) with a crystallinity of 18.9 wt.% in the swollen state can exceed 1,000 J/m$^2$. We further develop a strategy to enhance the anti-fatigue-fracture properties of PVA hydrogels, but still maintain their high water contents and low moduli by patterning highly-crystalline regions in the hydrogels. The current work not only reveals an anti-fatigue-fracture mechanism in hydrogels but also provides a practical method to design anti-fatigue-fracture hydrogels for diverse applications.

## Introduction

As polymer networks infiltrated with water, hydrogels have been widely used as scaffolds for tissue engineering (*1*), vehicles for drug delivery (*2*) and model platforms for biological studies (*3*). More recently, hydrogels have been explored for applications in devices and machines, including wearable electronics (*4, 5*), stretchable optical fibers (*6*), adhesives (*7-9*), soft robotics (*10*), and hydrogel-based soft machines (*11, 12*). The use of hydrogels in devices and machines requires them to maintain robustness under cyclic mechanical loads. Following the pioneering work by Gong et al (*13*), hydrogels have been made tough to resist crack propagation under a single cycle of mechanical load (*14-17*). The toughening of hydrogels is achieved by integrating mechanisms for dissipating mechanical energy such as fracture of short polymer chains and reversible crosslinks into stretchy polymer networks (*18, 19*). However, existing tough hydrogels still suffer from fatigue fracture under multiple cycles of mechanical loads (*11, 20, 21*), because the resistance to fatigue-crack propagation after prolonged cycles of loads is the energy required to fracture a single layer of polymer chains (i.e., the intrinsic fracture energy of the hydrogel), which is unaffected by the additional dissipation mechanisms introduced in tough hydrogels (*20, 21*). The reported fatigue thresholds of various tough hydrogels are 8.4 $J/m^2$ for polyacrylamide (PAAm)-polyvinyl alcohol (PVA) (*21*) and 53.2 $J/m^2$ for PAAm-alginate (*20*), the same order as their intrinsic fracture energies. The highest fatigue threshold for hydrogels reported so far is 418 $J/m^2$ for a double network hydrogel, poly(2-acrylamido-2-methyl-1-propanesulfonic acid) (PAMPS)-PAAm, which possibly can be attributed to the PAAm network with very long polymer chains and thus high intrinsic fracture energy (*13, 22*). A general strategy towards the design of anti-fatigue-fracture hydrogels has remained as a critical need and a central challenge for long-term applications of hydrogels in devices and machines.

In contrast to synthetic hydrogels, biological tissues such as cartilages, tendons, muscles and heart valves show extraordinary anti-fatigue properties. For example, the knee joint of an average person needs to sustain peak stresses of 4–9 MPa for 1 million cycles per year, and its fracture energy after prolonged cycles of loads is above 1,000 J/m$^2$ (*23, 24*). The anti-fatigue property of biological tissues possibly arises from the inherent highly ordered and partially crystalline structures of collagen fibers in the tissues (*25*). Inspired by anti-fatigue biological tissues, here we hypothesize that the increase of crystallinity in synthetic hydrogels can significantly enhance their fatigue thresholds, due to the need to fracture crystalline domains for fatigue-crack propagation (Fig. 1). The energy per unit area required to fracture crystalline domains of a polymer is much higher than that required to fracture a single layer of amorphous chains of the same polymer (*26*). To test the hypothesis, we select PVA as a model hydrogel with tunable crystallinity. We increase the annealing time (after freeze-thawing and air-drying) of the PVA hydrogel to give higher crystallinity, larger crystalline domain size, and smaller average distance between adjacent domains in the hydrogel (Fig. 1A). We then measure the fatigue thresholds of PVA hydrogels with various crystallinities (Fig. 1, B and C). We find that the increase of crystallinity can significantly enhance the fatigue thresholds of PVA hydrogels. In particular, the fatigue threshold can exceed 1,000 J/m$^2$ when the crystallinity of PVA in the swollen state reaches 18.9 wt.%. By annealing selected regions in hydrogels, we further demonstrate a general strategy to pattern highly-crystalline regions in the PVA hydrogels to render them resistant to fatigue-fracture but still maintain their high water content and low modulus. With this strategy, we create *kirigami hydrogel sheets* that are highly stretchable and resistant to fatigue-fracture by introducing patterned cuts into hydrogel sheets and then reinforcing the cut tips. The current work not only reveals a new anti-fatigue-fracture mechanism in hydrogels but also provides a practical method to design anti-fatigue-fracture hydrogels for diverse applications.

**Results**

To validate the hypothesis that the increase of crystallinity in hydrogels can significantly enhance their fatigue thresholds, we use PVA hydrogels as a model material system with tunable crystallinity. We first freeze a solution of uncrosslinked PVA at -20 °C for 8 h and thaw it at 25 °C for 3 h, to form a hydrogel crosslinked by crystalline domains (27). The freeze-thawed PVA hydrogel is further dried in an incubator at 37 °C and then annealed at 100 °C for various times ranging from 0 min to 90 min (28). (The hydrogel dry-annealed for 0 min means no annealing process.) The crystallinity of the hydrogel can be tuned by drying and annealing it for different times. As a control sample, we also fabricate a chemically-crosslinked PVA as a reference hydrogel composed of an entirely amorphous polymer network (See Materials and Methods for details on fabrication of the samples).

*Characterization of the crystalline morphology in PVA hydrogels*

We first measure the crystallinities of the resultant PVA hydrogels in their dry state by differential scanning calorimetry (DSC). For the chemically-crosslinked and freeze-thawed PVA hydrogels, we use excess chemical crosslinks to fix the amorphous polymer chains before air-drying them in an incubator at 37 °C (see Materials and Methods for details). The excess chemical crosslinks minimize the formation of further crystalline domains during the air-drying process (29, 30). As shown in Fig. 2A, both the chemically-crosslinked and freeze-thawed PVA hydrogels show negligible endothermic peaks with measured crystallinities in the dry state of 0.2 wt.% and 2.1 wt.%, respectively. However, when the freeze-thawed PVA (without excess chemical crosslinks) is dried in air, the crystallinity in the dry state increases to 37.7 wt.% (see Fig. 2A). The increase of the crystallinity implies that substantially more crystalline domains nucleate during the air-drying process (28). The crystallinity further increases gradually by increasing the annealing time. When the sample is annealed for 90 min at 100 °C, the crystallinity in the dry state reaches 47.3 wt.%. In addition to the crystallinities in the dry state measured from DSC, we further measure the water contents in the fully swollen samples shown in Fig. 2B and calculate the corresponding crystallinities in the swollen state shown in Fig. 2C. It is well-known that the as-prepared dry PVA samples may contain residual water bonded with polymer chains (30). The amount of residual water can be calculated from the endothermic transition ranging from 60 °C to 180 °C on the DSC curves (See Fig. 2A and fig. S1) (*27*). The above reported crystallinities in the dry state (without residual water), water contents and crystallinities in the swollen state have been corrected to account for the weights of residual water in the as-prepared dry samples (see Materials and Methods, table S1 and fig. S1 for details).

To quantify the evolution of crystalline morphology, we measure the average distance between adjacent crystalline domains $L$ through small-angle X-ray scattering (SAXS) and the average size of crystalline domains $D$ through wide-angle X-ray scattering (WAXS) (see details in Materials and Methods). We first perform SAXS measurements on the samples in the swollen state after subtracting the water background, measuring the scattering intensity $I(q)$ versus the scattering vector $q$. To identify the location of the peak intensity, we correct the intensity by multiplying the scattering intensity with the scattering vector $q$ (*31*). As shown in Fig. 2D, there is no peak in the plot of the corrected intensity $Iq$ versus the scattering vector $q$ for the freeze-thawed hydrogel, which implies negligible interference between adjacent crystalline domains. For the hydrogel dry-annealed for 0 min, there is a slight shoulder shown in the corrected intensity curve, which indicates stronger interference between adjacent domains. The average distance between adjacent crystalline domains $L$ can be estimated from the critical vector corresponding to the peak intensity $q_{max}$, following the Bragg expression $L = 2\pi / q_{max}$ (*31*). The average distance for the hydrogel dry-annealed for 0 min is estimated to be 21 nm in the swollen state. As the annealing time increases to 90 min, the average distance $L$ decreases to 13 nm in the swollen state (Fig. 2G). As a control case, we also measure SAXS profiles of the hydrogel dry-annealed 90 min in the dry state. As shown in Fig. 2F, the average distance between adjacent crystalline domains in the dry state is around 9 nm, smaller than the distance in the same hydrogel in the swollen state (i.e., 13 nm). This is because swelling of the interstitial amorphous polymer chains increases the distance between adjacent crystalline domains.

We further perform WAXS measurements on the hydrogels in their dry state using Ni-filtered CuKα1 radiation with X-ray wavelength $\lambda = 1.54$ Å. As shown in Fig. 2E, all dry-annealed PVA hydrogels show a strong diffraction peak at $2\theta = 19.7°$, which corresponds to the typical reflection plane of $(1\,0\,\bar{1})$ in semi-crystalline PVA (*32*). In addition, small peaks at $2\theta = 11.5°$ and $23.1°$ are also observed in the hydrogel dry-annealed for 90 min, suggesting a high crystallinity in the hydrogel, which is consistent with the DSC measurement. By identifying the half-width of the maximum diffraction peak $\beta$, the average size of crystalline domains $D$ can be approximately calculated using Scherrer's equation $D = k\lambda / (\beta \cos\theta)$ (*33*), where $k$ is a dimensionless shape factor varying with the actual shape of the crystalline domain; $\lambda$ is the wavelength of X-ray diffraction; and $\theta$ is the Bragg angle. Here, $\beta$ is identified after subtracting the instrumental line broadening; and the dimensionless shape factor $k$ is set as 1, approximating the spherical shape of the crystalline domains (*34*). As shown in Fig. 2G, by increasing the annealing time from 0 min to 90 min, the average size of the crystalline domains increases from 3.8 nm to 6.5 nm. This trend is consistent with the decrease of the average distance between adjacent crystalline domains with annealing time, since the growth of the crystalline domains consumes the interstitial amorphous polymer chains.

To further validate the tuning of crystalline domains in the PVA hydrogel with annealing time, we use tapping-mode atomic force microscopy (AFM) to obtain phase images of the hydrogels dry-annealed for 0 min and 90 min, respectively. The bright areas in Fig. 2H correspond to the regions with relatively high modulus (mainly crystalline domains), whereas the dark areas represent the regions with relatively low modulus (mainly amorphous domains). As shown in Fig. 2H, the morphology of isolated crystalline domains is observed in the hydrogel dry-annealed for 0 min, while the hydrogel dry-annealed for 90 min shows larger aggregated crystalline domains.

***Characterization of fatigue fracture properties of hydrogels***

To measure the fatigue threshold of PVA hydrogels, we adopt the single-notch method, which has been widely used in fatigue tests of rubbers (*35, 36*). Notably, all fatigue tests in this study are performed on fully-swollen hydrogels immersed in a water bath, to prevent the dehydration-induced crack propagation (fig. S2). We use dogbone-shaped samples and perform cyclic tensile tests on both notched and unnotched samples, which are otherwise the same (Fig. 1, B and C). The nominal stress versus stretch curves (i.e., $S$ vs. $\lambda$) of unnotched samples are obtained over $N$ cycles of applied stretch $\lambda^A$. The strain energy density $W$ of the unnotched sample under the $N_{th}$ cycle of applied stretch $\lambda^A$ can be calculated as

$$W(\lambda^A, N) = \int_1^{\lambda^A} S d\lambda \qquad (1)$$

where $S$ and $\lambda$ are the measured nominal stress and stretch, respectively. Thereafter, the same cyclic stretch $\lambda^A$ is applied on the notched sample, measuring the evolution of the cut length in undeformed state $c$ with the cycle number $N$. The applied energy release rate $G$ in the notched sample under the $N_{th}$ cycle of applied stretch $\lambda^A$ can be calculated as (*35, 36*)

$$G(\lambda^A, N) = 2k(\lambda^A) \cdot c(N) \cdot W(\lambda^A, N) \qquad (2)$$

where $k$ is a slowly varying function of the applied stretch expressed as $k = 3/\sqrt{\lambda^A}$, $c$ the current crack length at undeformed configuration, and $W$ the strain energy density measured in the unnotched sample (Eq. 1). By systematically varying the applied stretch $\lambda^A$, we can obtain a plot of crack extension per cycle versus the applied energy release rate (i.e., $dc/dN$ vs. $G$). By linearly extrapolating the curve of $dc/dN$ vs. $G$ to the intercept with the abscissa, we can approximately obtain the critical energy release rate $G_c$, below which the fatigue crack will not propogate under infinite cycles of loads. By definition, the fatigue threshold $\Gamma_0$ is equal to the critical energy release rate $G_c$. To validate that this extrapolated value $G_c$ is indeed the fatigue threshold $\Gamma_0$, we further apply $G_c$ to the notched sample over 30,000 cycles (to approximate infinite cycles of loads) and observe no crack extension (Fig. 3H). In addition, we also measure the fatigue thresholds using the pure-shear test (*20*) to validate the results from the single-notch test (fig. S6).

For cyclic tensile tests on unnotched samples (Fig. 3, A and B and fig. S3), both chemically-crosslinked PVA and freeze-thawed PVA show negligible Mullins effect, and their $S$ vs. $\lambda$ curves reach steady states after only a few cycles (i.e., 10 for chemically-crosslinked PVA at an applied stretch of 1.6, and 200 for freeze-thawed PVA at an applied stretch of 2.2). On the other hand, the dry-annealed PVA hydrogels exhibit a more significant Mullins effect, due to mechanical dissipation caused by melting and reorientation of crystalline domains (*31*). As the stretch further increases, the crystalline domains may transform into aligned fibrils along the loading direction (*31*). The energy required to damage the crystalline domains and fibrils is much higher than that to fracture a single layer of the same polymer in amorphous state. The hydrogel dry-annealed 90 min reaches steady state after 1,000 cycles of applied stretches of $\lambda^A = 2$ (Fig. 3C). Despite the Mullins effect, the steady-state maximum nominal stress of the hydrogel dry-annealed for 90 min is much higher than that of both the chemically-crosslinked PVA and the freeze-thawed PVA at the same applied stretch of $\lambda^A = 2$ (e.g., 2.3 MPa for hydrogel dry-annealed 90 min, 15 kPa for chemically-crosslinked PVA, and 3 kPa for freeze-thawed PVA in fig. S4).

For cyclic tensile tests on notched samples, a pre-crack is cut using a razor blade with tip radius of around 200 μm and initial crack length around 1 mm, smaller than one-fifth of the width of the sample (*36*). A digital microscope (AM4815ZT, Dino-Lite, resolution 20 μm/pixel) is used to record the cut length under cyclic loads (Materials and Methods, fig. S5). We first apply cyclic loads with a small applied stretch (i.e., $\lambda^A = 1.3$) on a notched sample. If the crack remains quasi-stationary with crack extension per cycle ($dc/dN$) smaller than 20 nm/cycle (i.e., no detectable crack extension in 1,000 cycles), the applied cyclic stretch is increased by the increment of $\Delta\lambda^A = 0.1$ for other notched samples until crack propagation greater than 20 nm/cycle is captured. As shown in Fig. 3, D and E, the fatigue thresholds of chemically-crosslinked PVA and freeze-thawed PVA are measured to be 10 J/m$^2$ and 23 J/m$^2$, respectively. The fatigue threshold for the hydrogel dry-annealed 0 min increases to 110 J/m$^2$. As the annealing time increases, the fatigue threshold further increases. In particular, for the hydrogel dry-annealed 90 min, the fatigue threshold can achieve 1000 J/m$^2$ (Fig. 3F and fig. S5). The measured fatigue threshold of the hydrogel dry-annealed 90 min from the pure-shear test is 918 J/m$^2$, consistent with the single-notch test (fig. S6). The dependence of fatigue threshold on the crystallinity is summarized in Fig. 3G. The fatigue theshold increases with the crystallinity and demonstrates a sharp jump when the crystallinity in the swollen state reaches approximate 15 wt.%.

In addition to fatigue tests, we also measure the nominal stress versus stretch curves of all hydrogels to obtain their Young's moduli and tensile strengths (fig. S7). As shown in Fig. 4, A and B, both the Young's modulus and tensile strength increase with the hydrogels' crystallinity and show dramatic enhancements when the crystallinity in the swollen state reaches approximate 15 wt.% (*37*). This sharp jump in Young's modulus and tensile strength is consistent with the dramatic increase of fatigue threshold of the hydrogel at approximate 15 wt.% crystallinity in the swollen state (Fig. 3G).

*Patterning of highly-crystalline regions in hydrogels*

Whereas annealing the whole PVA hydrogel can significantly enhance its fatigue threshold, the annealing treatment also increases the modulus and decreases the water content of the hydrogel (Fig. 4C and fig. S8). However, for many applications, it is desirable to maintain the relatively low modulus and high water content of the hydrogels. Here we propose a strategy to introduce programed highly-crystalline regions in the hydrogels. We use the computer-aided design of electrical circuits to induce localized heat treatment for annealing selected regions of the hydrogels (See details in Materials and Methods and fig. S9). The chemically-crosslinked PVA hydrogel is used as the pristine sample with low fatigue threshold of 15 J/m$^2$, low Young's modulus of 114 kPa and high water content of 88 wt.%. Three examples of programed annealing patterns on the pristine PVA hydrogels are demonstrated with enhanced fatigue thresholds.

We first locally introduce a highly-crystalline ring-shaped region around a crack tip (Fig. 5A), leading to a fatigue threshold over 236 J/m$^2$ (Fig. 5B). Despite the small area of the highly-crystalline region, this local conditioning can greatly reinforce the crack tip, delaying crack propagation. Meanwhile, the measured overall Young's modulus and water content of the sample are maintained at 114 kPa and 87 wt.% (Fig. 5, E and F), respectively, which are almost unaffected by the local annealing around the crack tip. As a second example, we pattern mesh-like highly-crystalline regions on the pristine hydrogel (Fig. 5C). Compared with the pristine sample, the fatigue threshhold of the mesh-reinforced sample increases to 290 J/m$^2$ (Fig. 5D), and its Young's modulus remains relatviely low (627 kPa) and its water content stays relatively high as 83 wt.% (Fig. 5, E and F).

The strategy of patterning highly-crystalline regions can be applied to various structures of hydrogels for improving anti-fatigue performances as well. For example, the kirigami structure is commonly adopted to enhance the stretchablity (*38*) and effective adhesion (*39*) of films. As a third example, we demonstrate a kirigami hydrogel sheet with improved anti-fatigue performance by patterning highly-crystalline regions around the cut tips. As shown in Fig. 5G, we first generate a kirigami pattern on a pristine sample by introducing parallel periodic cuts with equal length and equal distance bewteen adjacent cuts. The ultimate stretch and effective nominal stress (i.e., force divided by cross-section area) of the pristine kirigami hydrogel sheet under a single cycle of load are measured to be 1.7 and 0.9 kPa, respectively. Thereafter, the pristine kirigami sheet is reinforced around the cut tips with local annealing. The reinforced kirigami sheet can sustain cyclic tensile loads at an applied stretch of $\lambda^A = 2.1$ without detectable fatigue-fracture even when the cycle number approaches 3,000. Moreover, the effective nominal stress after prolonged cycles at the applied stretch of $\lambda^A = 2.1$ can maintain the plateau of 13 kPa, which is 14.4 times the strength of the pristine kirigami sheet (Fig.5, H and I).

Figure 5J and K compare the fatigue thresholds, water contents and the Young's moduli of reported hydrogels in literature (*20-22, 40, 41*). We show that by patterning highly-crystalline regions, both tip-reinforced and mesh-reinforced PVA hydrogels outperform existing synthetic hydrogels in terms of fatigue thresholds, and they can still maintain relatively high water contents and relatively low Young's moduli.

## Discussion

We have demonstrated that the fatigue threshold of hydrogels can be greatly enhanced by designing crystalline domains in the hydrogels. We use PVA hydrogels as a model material to validate this new mechanism for designing anti-fatigue-fracture hydrogels. The fatigue threshold of a PVA hydrogel with a crystallinity of 18.9 wt.% in the swollen state can achieve over 1000 J/m$^2$. We further develop a strategy to make PVA hydrogels anti-fatigue-fracture but still maintain their high water content and low modulus by patterning highly-crystalline regions in the hydrogels. The reported mechanism and strategy for designing anti-fatigue-fracture hydrogels can be extended to hydrogel composites with fillers such as nanoclay, nanocellulose and nanofibers.

The capability to enhance the anti-fatigue-fracture performance of synthetic hydrogels makes a number of future research directions and applications possible. For example, anti-fatigue hydrogels can be used for hydrogel-based gastric-retentive devices and implantable tissue replacements of meniscus, intervertebral disk and cartilage, which require long-term mechanical robustness when interacting with the human body.

## Materials and Methods

### Synthesis of PVA hydrogels

All types of our PVA hydrogels are synthesized from 10 wt.% poly(vinyl alcohol) (PVA; Mw 146,000-186,000, 99+% hydrolyzed; Sigma-Aldrich, 363065) solution. The solution was heated in a water bath at 100 °C with stirring for 5 h. To synthesize chemically-crosslinked PVA hydrogels, we added 10 µL glutaraldehyde (25 vol.%, Sigma-Aldrich, G6257) as a crosslinker to 1 mL 10 wt.% PVA solution, and added 10 µL hydrochloric acid (36.5-38 wt.%, J.T. Baker, 9535-02) as an accelerator into the other 1 mL 10 wt.% PVA solution. We then mixed and defoamed each of them by using a centrifugal mixer (AR-100; Thinky). The final mixtures, obtained by mixing and defoaming the two solutions together, were then cast into a mold and allowed to cure for 2 h. The chemically-crosslinked PVA hydrogels were immersed in deionized water for two days to remove unreacted chemicals. To fabricate freeze-thawed PVA hydrogels, 10 wt.% PVA solutions after mixing and defoaming were poured into a mold, frozen at -20 °C for 8 h and thawed at 25 °C for 3 h. The freeze-thawed hydrogels were further dried in an incubator (New Brunswick Scientific, C25) at 37 °C for 2 h, and then annealed at 100 °C for a controlled time (i.e., 0, 1, 3, 5, 10, or 90 min). All as-prepared PVA hydrogels were immersed in water to achieve their equilibrium-swollen state.

**Measurement of residual water and crystallinity in dry samples**

We measured the crystallinities of the resultant PVA hydrogels in their dry state by differential scanning calorimetry (DSC/cell: RCS1-3277 Cooling System: DSC1-0107). For as-prepared chemically-crosslinked PVA and freeze-thawed PVA, we used excess chemical crosslinks to fix the amorphous polymer chains before air-drying, minimizing the formation of further crystalline domains during the air-drying process. We first soaked the samples (thickness of 1mm) in the aqueous solution consisting of 10 mL glutaraldehyde (25 vol.%, Sigma-Aldrich, G6257), 500 µL hydrochloric acid (36.5-38 wt.%, J.T. Baker, 9535-02) and 50 mL DI water for 2 h. Thereafter, we soaked the samples in a DI water bath for 2 h to remove the residual hydrochloric acid. The samples were further dried in an incubator (New Brunswick Scientific, C25) at 37 °C for 2 h.

Thereafter, we measured the mass of residual water $m_{residual}$, the mass of crystalline domains $m_{crystalline}$ and the total mass of the dry samples (with residual water) $m$ using DSC. In a typical DSC measurement, we first weighed the total mass of the dry sample (with residual water) $m$. The sample were thereafter placed in a Tzero-pan and heated up from 50 °C to 250 °C at the rate of 20 °C/min under a nitrogen atmosphere with flow rate of 30 mL/min. The curve of heat flow shows a broad peak from 60 °C to 180 °C, indicating that the sample contains a small amount of residual water. The integration of the endothermic transition ranging from 60 °C to 180 °C gives the enthalpy for evaporation of the residual water per unit mass of the dry sample (with residual water) $H_{residual}$. Therefore, the mass of the residual water $m_{residual}$ can be calculated as

$$m_{residual} = m \cdot \frac{H_{residual}}{H^0_{water}} \quad (3)$$

where $H^0_{water} = 2260$ J/g is the latent heat of water evaporation. The curve of heat flow shows another narrow peak ranging from 200 °C to 250 °C corresponding to melting of the crystalline domains. The integration of the endothermic transition ranging from 200 °C to 250 °C gives the enthalpy for melting the crystalline domains per unit mass of the dry sample (with residual water) $H_{crystalline}$. Therefore, the mass of the crystalline domains $m_{crystalline}$ can be calculated as

$$m_{crystalline} = m \cdot \frac{H_{crystalline}}{H^0_{crystalline}} \quad (4)$$

where $H^0_{crystalline} = 138.6$ J/g is the enthalpy of fusion of 100 wt.% crystalline PVA measured at the equilibrium melting point $T^0_m$ (*30*). Therefore, the crystallinity in the ideally dry sample $X_{dry}$ (without residual water) can be calculated as

$$X_{dry} = \frac{m_{crystalline}}{m - m_{residual}} \quad (5)$$

**Measurement of water content and crystallinity in swollen samples**

The swollen hydrogels weighing $m_{swollen}$ were placed in an incubator (New Brunswick Scientific, C25) at 37 °C for 2 h, and weighed $m$ after air-drying. The mass of the residual water in the as-prepared dry samples $m_{residual}$ was measured in the previous section. Therefore, the water content in the swollen state can be calculated as $(m_{swollen} - m + m_{residual})/m_{swollen}$ and the polymer content in the swollen state can be calculated as $(m - m_{residual})/m_{swollen}$. In addition, the crystallinity in the swollen sample $X_{swollen}$ can be calculated as

$$X_{swollen} = \frac{m_{crystalline}}{m_{swollen}} = X_{dry} \cdot \frac{m - m_{residual}}{m_{swollen}} \tag{6}$$

where $m_{crystalline}$ and $X_{dry}$ were measured as described in the previous section.

**AFM phase imaging**

AFM phase images were acquired by atomic force microscope (MFP-3D, Asylum Research) in tapping mode. Dry free-standing PVA films were directly attached onto sample stage by double-sided carbon tape. The probe lightly taps on the sample surface with recorded phase shift angle of the probe motion relative to a driving oscillator. The bright regions with high phase angle correspond to regions with relatively high modulus; and the dark regions with low phase angle correspond to regions with relatively low modulus.

**X-ray scattering**

The X-ray scattering measurement was performed with Pilatus3R 300K detector Bruker Nanostar SAXS in X-ray Diffraction Shared Experimental Facility at MIT. We used small angle 2 mm beamstop with sample-detector distance of 1059.1 mm for SAXS measurements and wide angle 2mm beamstop with sample-detector distance of 109.1 mm for WAXS measurements. The exposure time was set as 300 s. Raw SAXS and WAXS patterns were processed with corrections by MATLAB-based GIXSGUI software before analysis.

**Raman spectroscopy**

Each sample was hydrated for more than 2 h and then pressed between a glass slide and a cover slip in order to ensure a flat surface. The cover slip was then sealed at the edges with nail polish to prevent the hydrogel from drying. A confocal Raman microscope (Alpha300RA; WiTec, Germany) with 20× objective (Zeiss, Germany) was used. Nd:YAG laser (532 nm) was used as the excitation source with the maximum power of 75mW. Data was collected with a CCD detector (DU401A-BV; Andor, UK) behind a 600 g/mm grating spectrometer (UHTS 300; WiTec, Germany). A 20-μm resolution Raman map of 4×3 mm scan area was acquired with an accumulation time of 1 second per point. Each point was pre-bleached for 400 ms to decrease the effect of fluorescence. Cosmic ray removal and background subtraction were performed to clean the spectra. The intensity of O-H bond within the PVA and water were calculated by integrating the spectra in the range of 2,800-3,000 cm$^{-1}$ and 3,075-3,625 cm$^{-1}$, respectively. The ratio of PVA and water was then calculated and plotted as a heatmap shown in Fig. 5A.

**Measurement of the fatigue threshold**

All the mechanical tests were performed in a water bath at 25°C with a U-stretch testing device (CellScale, Canada). For mechanically weak samples (e.g., the Ch and FT hydrogels), a load cell with maximum force of 4.4 N was used; for mechanically strong samples (e.g., hydrogel dry-annealed 90 min), a load cell with maximum force of 44 N was used. The dogbone-shaped sample had the dimensions at as-prepared state with the width of 5mm, thickness of 0.8mm and gauge length of 10mm. The nominal stress $S$ was measured from the recorded force $F$ divided by width $W$ and thickness $t$ in the swollen state. To measure the applied stretch $\lambda^A$ in gauge length, we adopted the digital image correlation method and calibrated the correlation between $\lambda^A$ and loading distance $d$ at gripping points (fig. S2). A digital microscope (AM4815ZT; Dino-Lite, resolution 20 μm/pixel) was used to record the crack extension. Since dry annealed PVA is transparent when immersing in water, we spread a small amount of graphite powder on the surface of the sample for visualization.

**Preparation of samples with programmable crystalline domains**

A copper tape was first loaded in a paper cutting machine (Silhouette CAMEO 3). The electrical circuits were first designed with CAD and loaded to the machine via Silhouette Studio software. After completion of the cutting, the copper tape was placed on a flat acrylic plate and unwanted parts were carefully peeled off. Two additional conductive wires were soldered on the copper tape as two electrodes. An electrical generator (Instek PSB-2400L2; Tequipment) was used to apply controlled current around 2-5 A on the designed circuits. To ensure achieving the designed temperature, a thermal imager (Seek thermal XR imager) was first used to measure the local temperature from the circuit (see fig. S9). By adjusting the current to control the designed temperature, the air-dried PVA hydrogel was placed on the targeted circuits for local heating. A gripper was used to ensure fully contact between the PVA film and copper circuits.

**Supplementary Material**

table S1. Crystallinities and water contents in chemically-crosslinked (Ch), freeze-thawed (FT) and dry-annealed PVA with annealing time of 0, 1, 3, 5, 10, 90 min.

fig. S1. Measurement of the mass of freeze-thawed PVA during air-drying and the amount of residual water in the sample after air-drying.

fig. S2. Experiment method for measuring fatigue threshold.

fig. S3. Shakedown softening of three types of PVA hydrogels.

fig. S4. Steady-state nominal stress versus stretch curves of PVA hydrogels with various crystallinities.

fig. S5. Validation of high fatigue threshold with single-notch test.

fig. S6. Validation of high fatigue threshold with pure shear test.

fig. S7. Mechanical characterization of PVA hydrogels with various crystallinities.

fig. S8. Raman spectroscopy of freeze-thawed PVA and 90-min dry-annealed PVA.

fig. S9. Electrical circuit and thermal mapping for programmable crystalline domains.

movie S1. Tension of the pristine notched sample and the tip-reinforced sample.

movie S2. Cyclic loading of the reinforced kirigami hydrogel sheet.

**Acknowledgments**

The authors thank Dr. A. F. Schwartzman in the Nano Mechanical Technology Laboratory at MIT for his help with AFM phase imaging, J. Zhou at MIT for discussion on DSC measurements and Z. Jiang at Argonne National Laboratory for discussion on SAXS results. **Funding:** This work is supported by National Science Foundation (CMMI-1661627), Office of Naval Research (N00014-17-1-2920), National Institutes of Health (EB000244), and MIT Institute for Soldier Nanotechnologies (CMMI-1253495). H.Y. acknowledges the financial support from Samsung Scholarship. **Author Contributions:** S.L. and X.Z. conceived the idea, designed the study and interpreted the results. S.L. performed DSC, SAXS and AFM measurements. S.L. and X.L. performed mechanical characterization and fatigue tests. S.L., C.S. and J. S. interpreted the SAXS results. H.L. performed Raman spectroscopy. S.L., X.L., H.Y. and J.L. performed the demonstration of patterning highly crystalline regions. S.L., X.L., H.Y., J.L., G.A.P, C.S., G.H.M., and X.Z. analyzed and interpreted the results. S.L. and X.Z. drafted the manuscript with inputs from all other authors. X.Z. supervised the study. **Competing interests:** The authors declare that they have no competing interests. **Data and materials availability:** The data that support the findings of this study are available from the corresponding author upon request.

# Figures and Tables

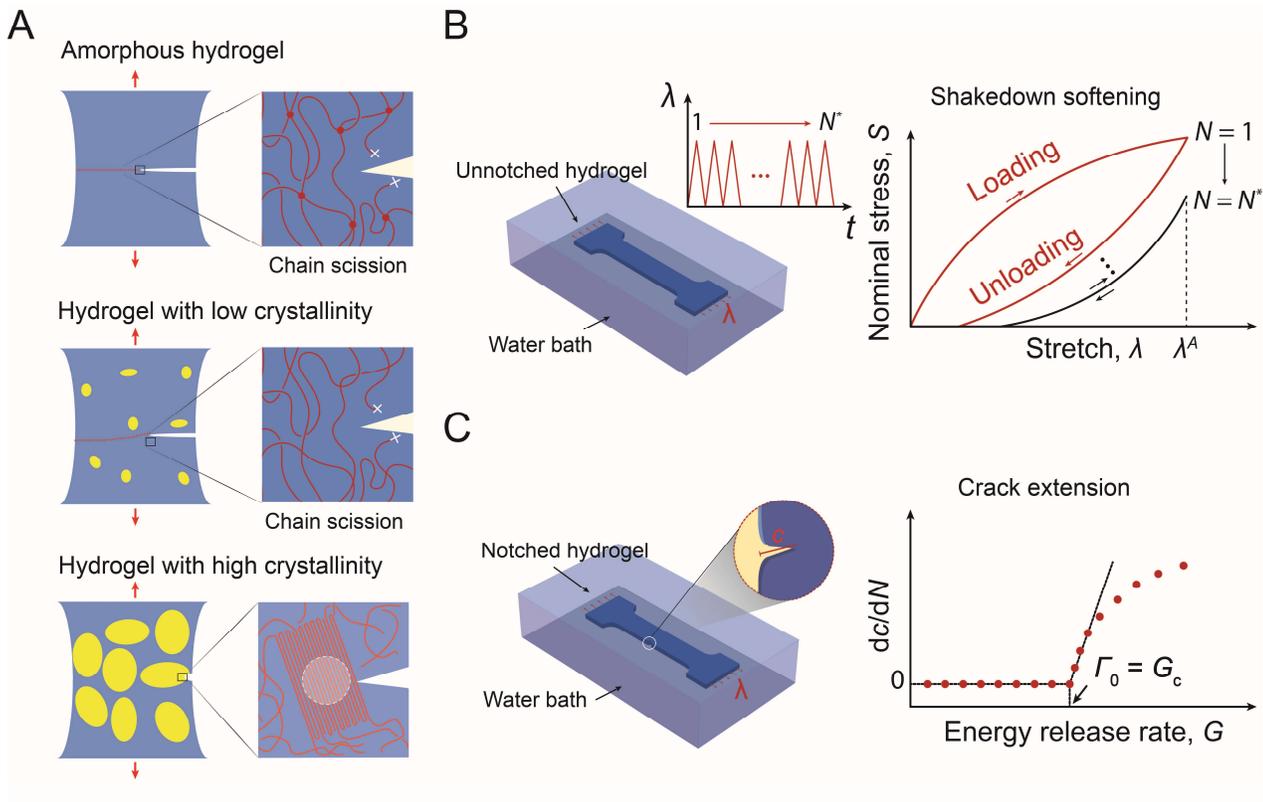

**Fig. 1. The design strategy for anti-fatigue-fracture hydrogels.** (**A**) Illustration of fatigue-crack propagation in an amorphous hydrogel and in hydrogels with low and high crystallinities under cyclic loads. The yellow areas represent crystalline domains and the blue areas denote amorphous domains. In the amorphous hydrogel and the hydrogel with low crystallinity, the fatigue threshold can be attributed to the energy required to fracture a single layer of polymer chains per unit area. In the hydrogel with high crystallinity, the fatigue-crack propagation requires fracture of crystalline domains. (**B**) Illustration of measuring nominal stress $S$ vs. stretch $\lambda$ curves over $N$ cycles of the applied stretch $\lambda^A$. The stress-stretch curve reaches steady state as $N$ reaches a critical value $N^*$. (**C**) Illustration of measuring crack extension per cycle $dc/dN$ vs. energy release rate $G$ curves. By linearly extrapolating the curve to intercept with the abscissa, we can approximately obtain the critical energy release rate $G_c$, below which the fatigue crack will not propagate under infinite cycles of loads. By definition, the fatigue threshold $\Gamma_0$ is equal to the critical energy release rate $G_c$.

Page 21 of 26

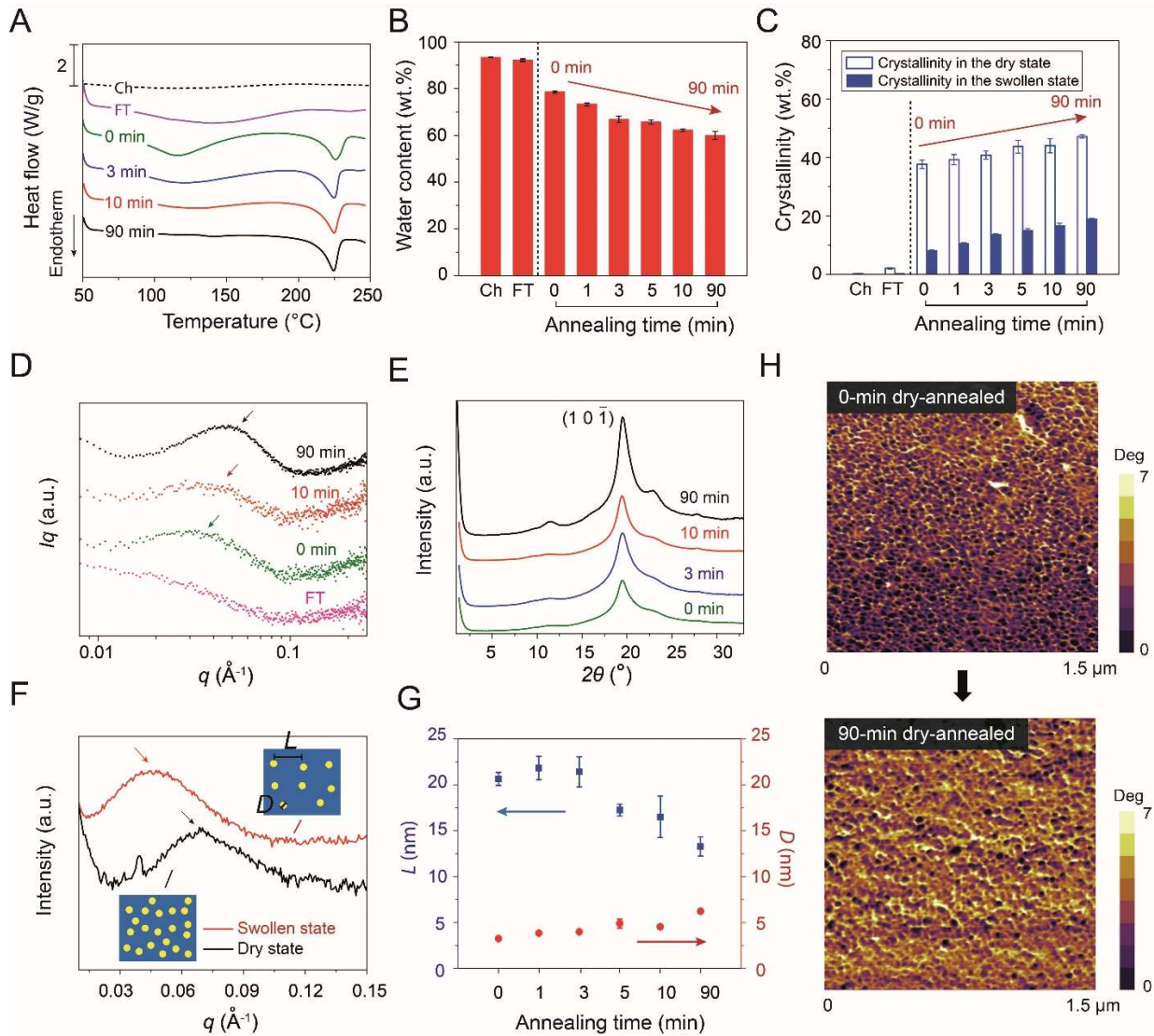

**Fig. 2. Characterization of crystalline domains in PVA hydrogels.** (**A**) Representative DSC thermographs of chemically-crosslinked (Ch), freeze-thawed (FT) and dry-annealed PVA with annealing time of 0, 3, 10, 90 min. (**B**) Water contents of chemically-crosslinked (Ch), freeze-thawed (FT) and dry-annealed PVA with annealing time of 0, 1, 3, 5, 10, 90 min. (**C**) Measured crystallinity in the dry and swollen states of Ch, FT and dry-annealed PVA with annealing time of 0, 1, 3, 5, 10, 90 min. (**D**) Representative SAXS profiles of FT hydrogel and dry-annealed PVA with annealing time of 0, 10, and 90 min. (**E**) Representative WAXS profiles of annealed PVA with annealing time of 0, 3, 10, and 90 min. (**F**) SAXS profiles of 90-min dry-annealed PVA in the dry state and the swollen state. The insets illustrate the increase of the distance between adjacent crystalline domains due to swelling of amorphous polymer chains. (**G**) The estimated average distance between adjacent crystalline domains $L$ and average crystalline domain size $D$ of dry-annealed PVA with annealing time of 0, 1, 3, 5, 10, 90 min. (**H**) AFM phase images of dry-annealed hydrogel with annealing time of 0 min and 90 min. Data in **B**, **C** and **G** are means ± SD, n = 3.

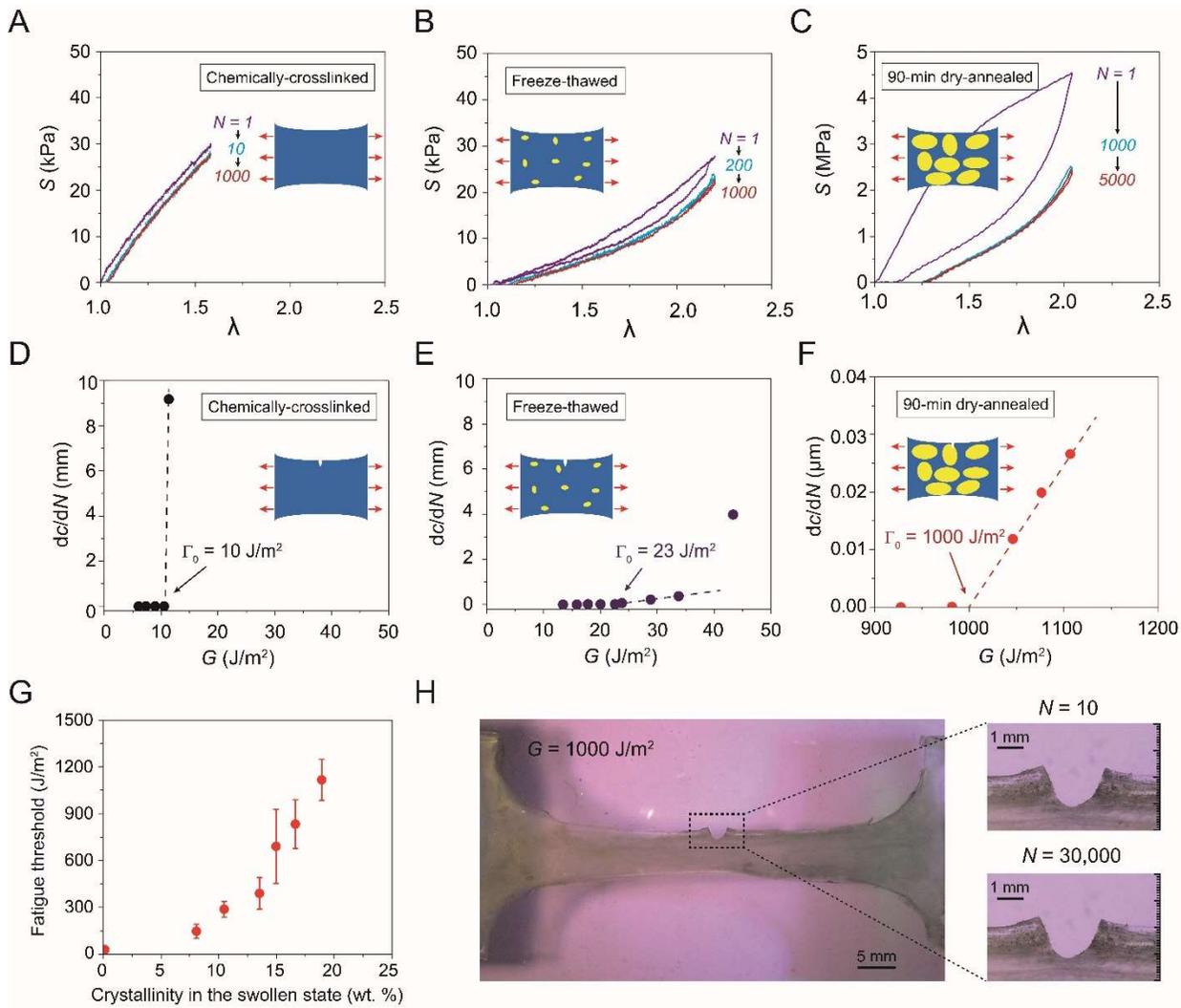

**Fig. 3. Measurement of fatigue thresholds of PVA hydrogels.** Nominal stress $S$ vs. stretch $\lambda$ curves over cyclic loads for (**A**) chemically-crosslinked hydrogel at an applied stretch of $\lambda^A = 1.6$, (**B**) freeze-thawed hydrogel at an applied stretch of $\lambda^A = 2.2$, and (**C**) 90-min dry-annealed hydrogel at an applied stretch of $\lambda^A = 2.0$. Crack extension per cycle $dc/dN$ vs. applied energy release rate $G$ for (**D**) chemically-crosslinked hydrogel, (**E**) freeze-thawed hydrogel, and (**F**) dry-annealed hydrogel with annealing time of 90 min. (**G**) The fatigue threshold increases with the crystallinity of the hydrogel in the swollen state. (**H**) Validation of fatigue threshold as high as 1000 J/m$^2$ in 90-min dry-annealed hydrogel using the single-notch test. Data in **G** are means ± SD, n = 3. Scale bars are 5 mm and 1 mm for left and right images in **H**.

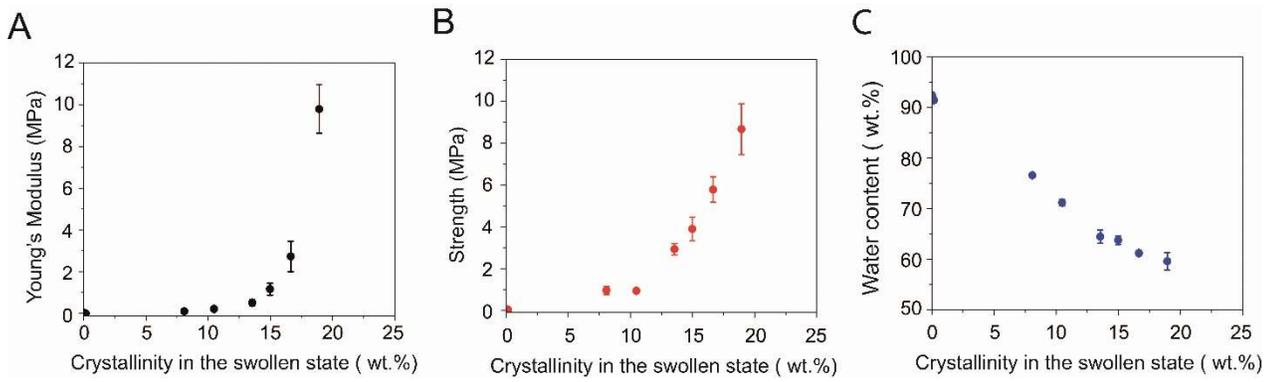

**Fig. 4**. **Young's moduli, tensile strengths and water contents of PVA hydrogels.** (**A**) Young's modulus versus crystallinity in the swollen state. (**B**) Tensile strength versus crystallinity in the swollen state. (**C**) Water content versus crystallinity in the swollen state. Data in **A**, **B** and **C** are means ± SD, n = 3.

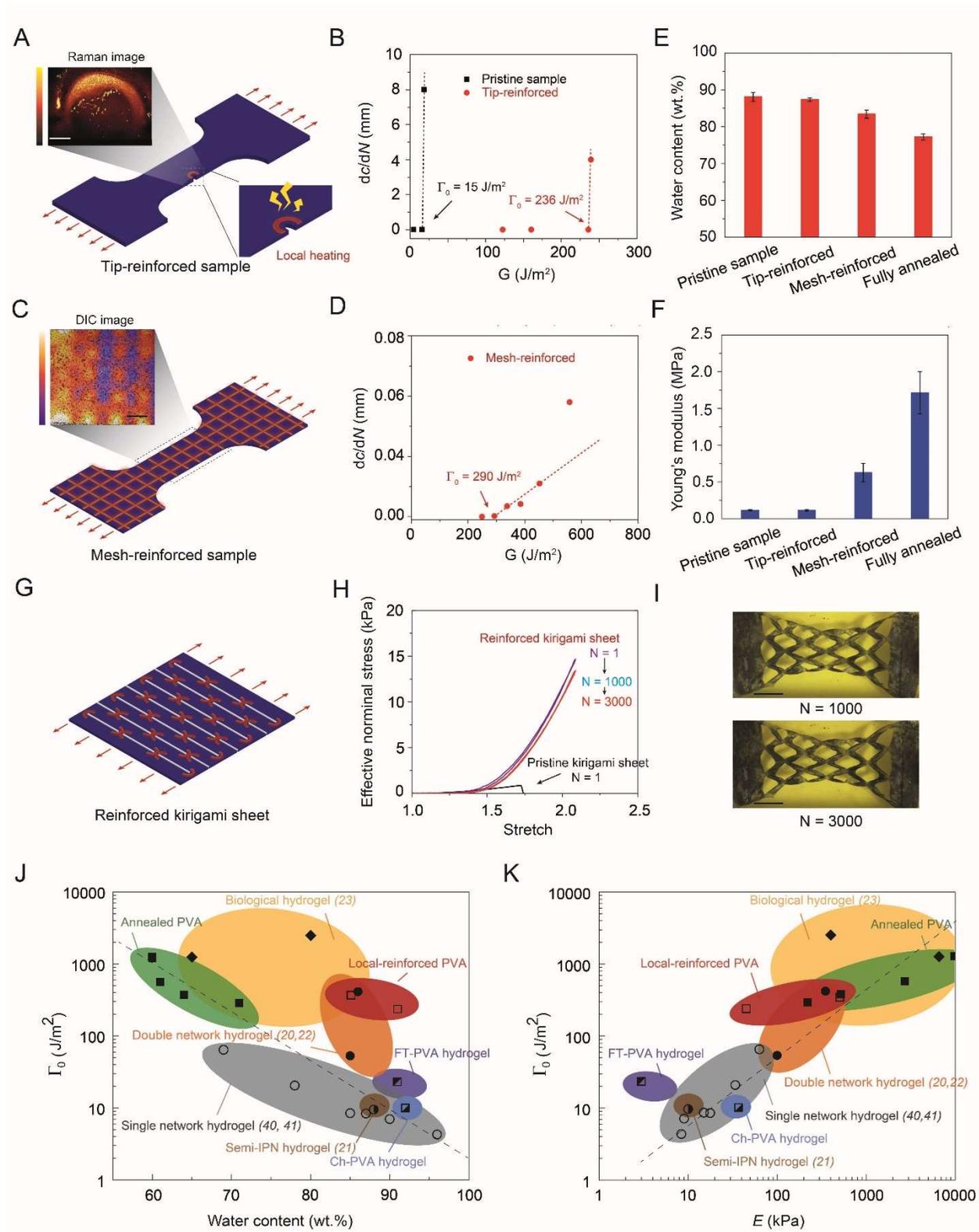

**Fig. 5**. **Patterning highly-crystalline regions in PVA hydrogels.** (**A**) Illustration of introducing a highly-crystalline region around crack tip. Inset: Raman spectroscopy with bright color representing low water content and dark color representing high water content (see details in Materials and Methods). (**B**) Comparison of crack extension per cycle d$c$/d$N$ vs. applied energy releasing rate $G$ between the pristine sample and the tip-reinforced sample. The fatigue thresholds of the pristine sample and the tip-reinforced sample are 15 J/m$^2$ and 236 J/m$^2$, respectively. (**C**) Illustration of introducing mesh-like highly-crystalline regions. Inset: Digital image correlation method (DIC) shows large deformation in low-crystallinity regions and small deformation in high-crystallinity regions. (**D**) Crack extension per cycle d$c$/d$N$ vs. applied energy releasing rate $G$ of the mesh-reinforced sample. The fatigue threshold of the mesh-reinforced sample is 290 J/m$^2$. (**E**) Water contents of the pristine sample, the tip-reinforced sample, the mesh-reinforced sample and the fully-annealed PVA hydrogels. (**F**) Young's moduli of the pristine sample, the tip-reinforced sample, the mesh-reinforced sample and the fully-annealed PVA hydrogels. (**G**) Illustration of introducing highly-crystalline regions around cut tips in a pristine kirigami sheet. (**H**) Effective nominal stress vs. stretch curves of the reinforced kirigami sheet under cyclic loading. Effective nominal stress vs. stretch curve of the pristine kirigami sheet under a single cycle of load. (**I**) Images of the reinforced kirigami sheet under 1000$^{th}$ cycle and under 3000$^{th}$ cycle. (**J**) Comparison of fatigue thresholds and water contents among reported synthetic hydrogels (*20-22, 40, 41*), PVA hydrogels with patterned highly-crystalline regions and biological tissues (*23*). (**K**) Comparison of fatigue thresholds and Young's moduli among reported synthetic hydrogels, PVA hydrogels with patterned highly-crystalline regions and biological tissues. Data in **E** and **F** are means ± SD, n = 3. Scale bars are 800 μm for **A**, 1 mm for **C**, and 40 mm for **I**.

**This PDF file includes:**

Table S1. Crystallinities and water contents in chemically-crosslinked (Ch), freeze-thawed (FT) and dry-annealed PVA with annealing time of 0, 1, 3, 5, 10, 90 min.

Fig. S1. Measurement of the mass of freeze-thawed PVA during air-drying and the amount of residual water in the sample after air-drying.

Fig. S2. Experiment method for measuring fatigue threshold.

Fig. S3. Shakedown softening of three types of PVA hydrogels.

Fig. S4. Steady-state nominal stress versus stretch curves of PVA hydrogels with various crystallinities.

Fig. S5. Validation of high fatigue threshold with single-notch test.

Fig. S6. Validation of high fatigue threshold with pure shear test.

Fig. S7. Mechanical characterization of PVA hydrogels with various crystallinities.

Fig. S8. Raman spectroscopy of freeze-thawed PVA and 90-min dry-annealed PVA.

Fig. S9. Electrical circuit and thermal mapping for programmable crystalline domains.

Legend for Movie S1.

Legend for Movie S2.

| Sample name | Crystallinity in the dry state without residual water (wt. %) | Water content in the swollen state (wt. %) | Crystallinity in the swollen state (wt. %) |
|---|---|---|---|
| Ch | 0.23±0.16 | 93.4±0.2 | 0.02±0.01 |
| FT | 2.1±0.2 | 92.2±0.5 | 0.16±0.02 |
| 0 min | 37.7±1.4 | 78.6±0.4 | 8.1±0.3 |
| 1 min | 39.3±1.7 | 73.3±3.5 | 10.5±0.5 |
| 3 min | 40.9±1.4 | 66.9±1.3 | 13.5±0.5 |
| 5 min | 43.8±2.2 | 65.8±0.9 | 15.0±0.8 |
| 10 min | 44.1±2.4 | 62.2±2.6 | 16.6±0.9 |
| 90 min | 47.3±0.6 | 59.9±1.7 | 18.9±0.2 |

**Table S1.** Crystallinities and water contents in chemically-crosslinked (Ch), freeze-thawed (FT) and dry-annealed PVA with annealing time of 0, 1, 3, 5, 10, 90 min.

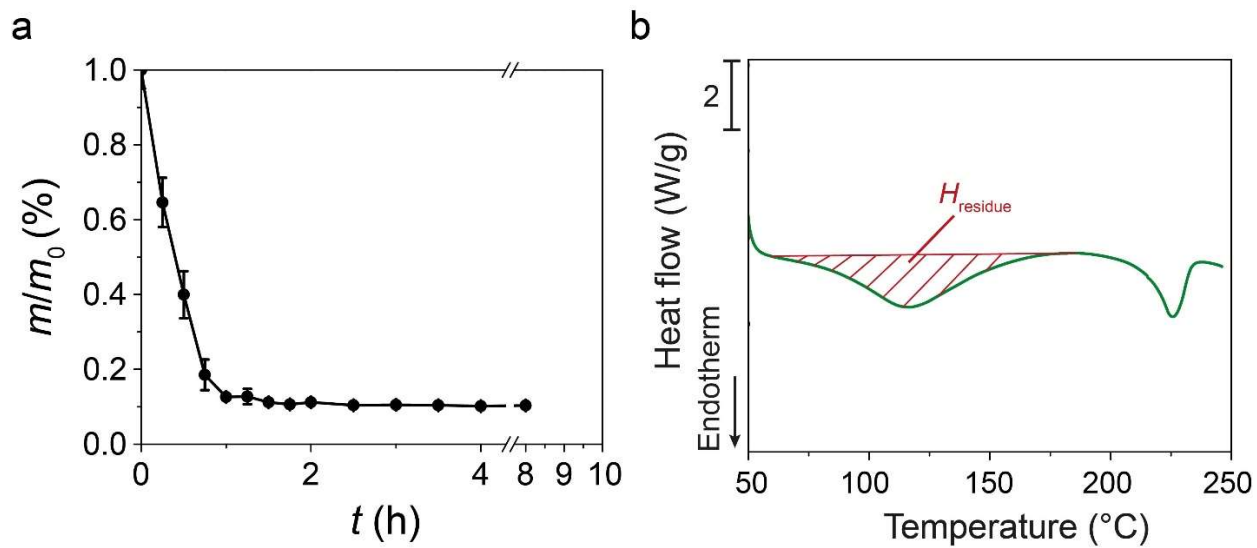

**Fig. S1**. **a**. Measured mass normalized by the initial mass (at swollen state) of freeze-thawed PVA sample versus air-drying time. **b**. The amount of residual water can be calculated from the endothermic transition ranging from 60 °C to 180 °C on the DSC curves. (See materials and methods for details.)

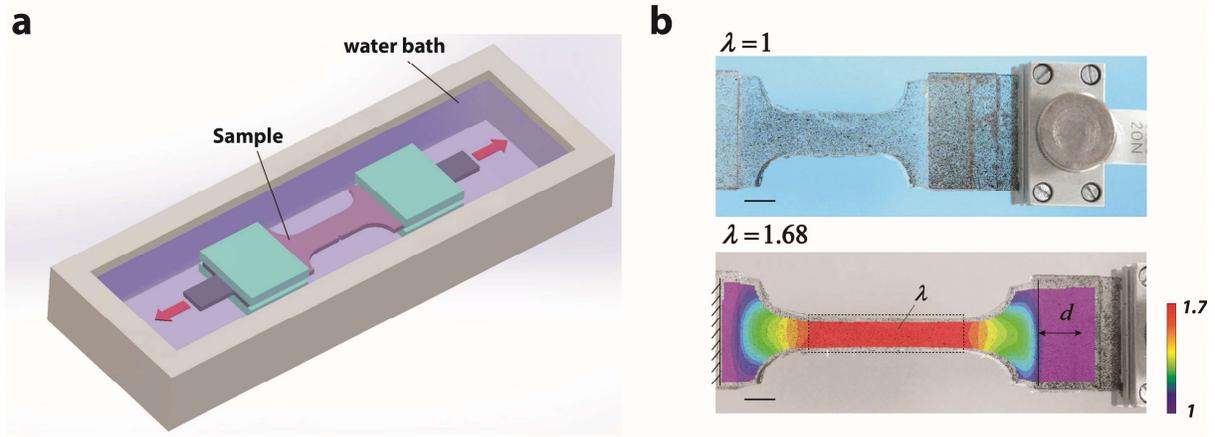

**Fig. S2. Experiment method for measuring fatigue threshold. a**, Schematic illustration of test setup. **b**, Digital correlation method for calibration of the correlation between stretch $\lambda$ in gauge length and applied distance $d$. Scale bars in **b** are 5 mm.

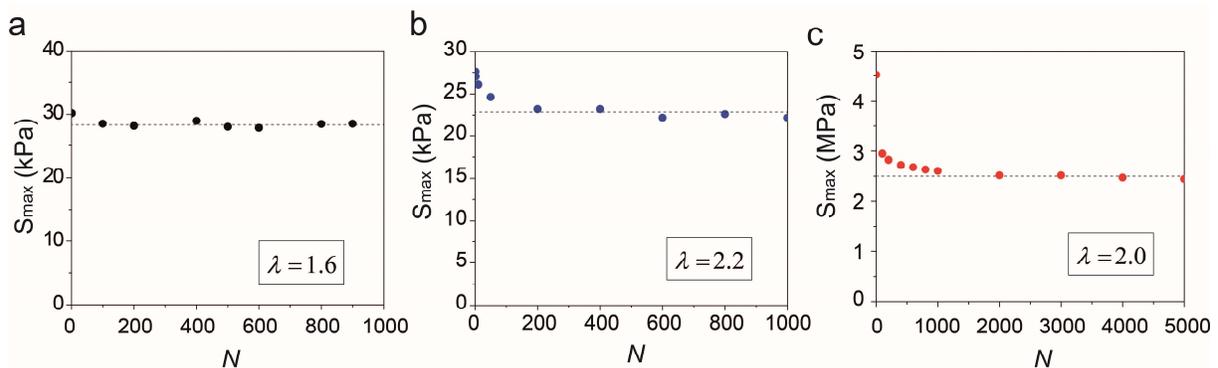

**Fig. S3**. **Shakedown** softening of three types of PVA hydrogels. **a**, Maximum stress at the stretch of 1.6 versus cycle number of chemically-crosslinked hydrogel, **b**, Maximum stress at the stretch of 2.2 versus cycle number of freeze-thawed PVA and **c**, Maximum stress at the stretch of 2.0 versus cycle number of 90-min dry-annealed PVA.

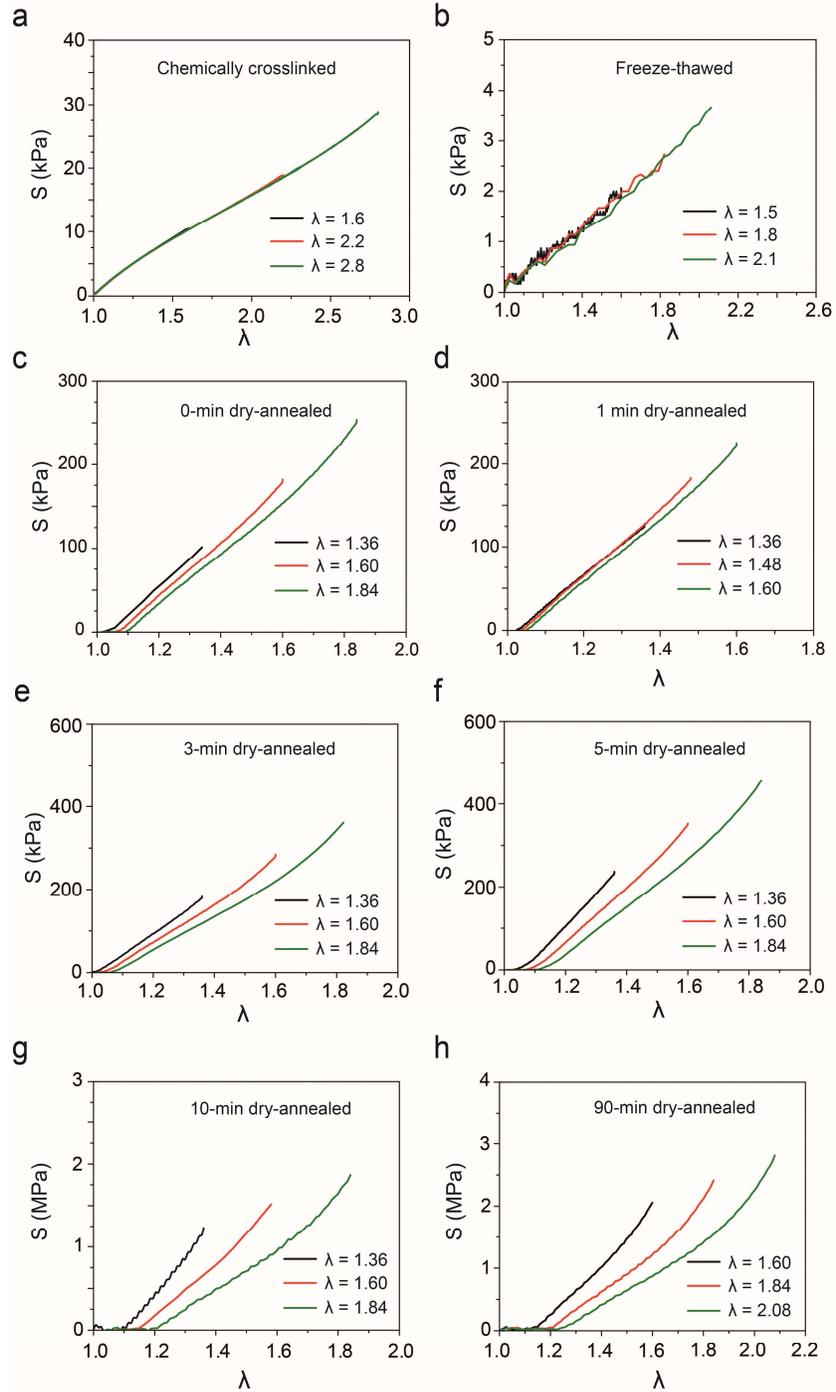

**Fig. S4**. **Steady-state nominal stress versus stretch** curves of PVA hydrogels with various crystallinities. **a**, Chemically-crosslinked PVA, **b**, Freeze-thawed PVA, **c**, 0-min dry-annealed PVA, **d**, 1-min dry-annealed PVA, **e**, 3-min dry-annealed PVA, **f**, 5-min dry-annealed PVA, **g**, 10-min dry-annealed PVA and **h**, 90-min dry-annealed PVA.

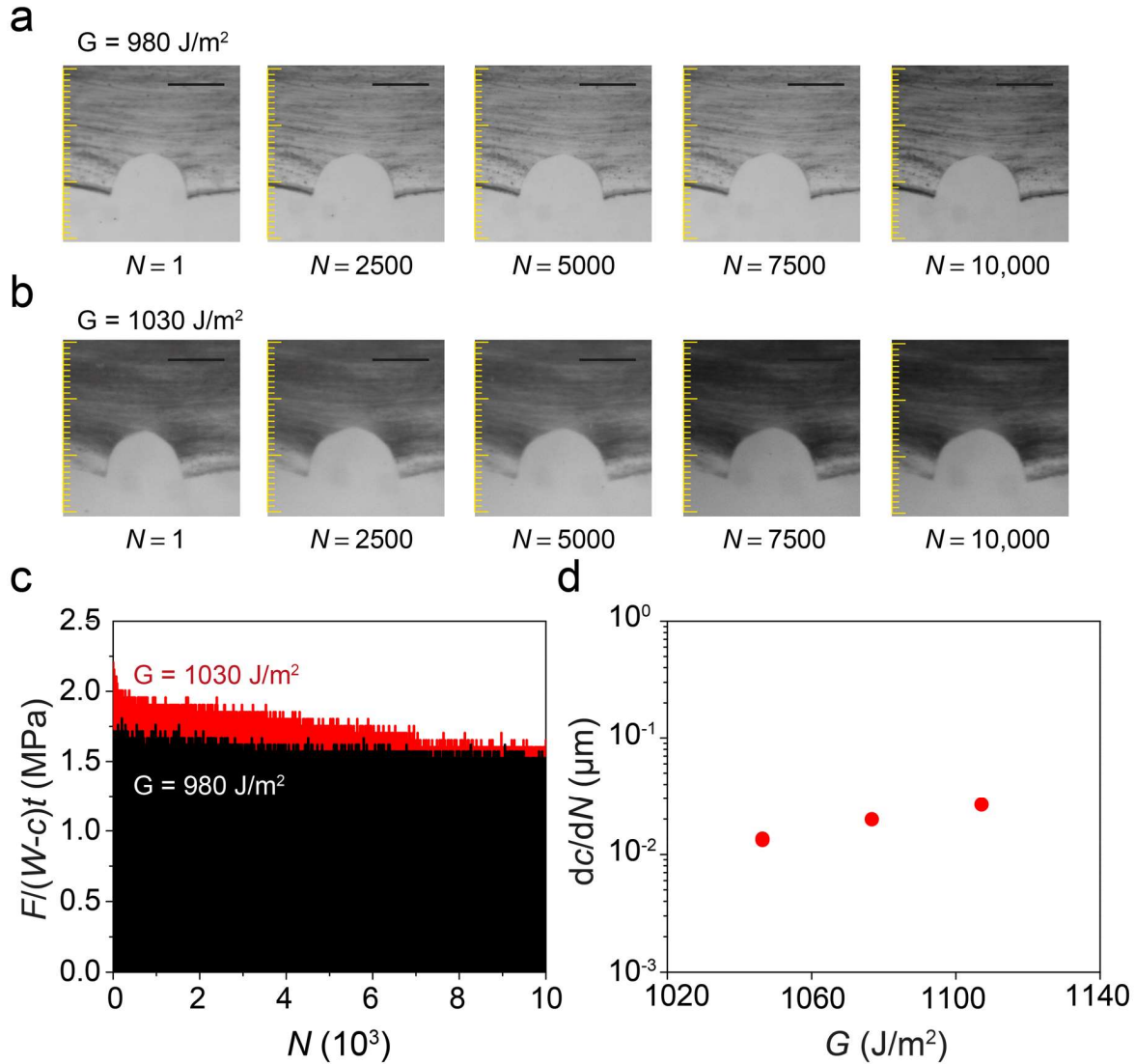

**Fig. S5. Validation of high fatigue threshold with single-notch test.** The material is 90-min dry-annealed PVA. **a**, The images of deformed sample under the initial energy release rate of 980 J/m$^2$ at the cycle numbers of 1, 2500, 5000, 7500 and 10,000. **b**, The images of deformed sample under the initial energy release rate of 1030 J/m$^2$ at the cycle numbers of 1, 2500, 5000, 7500 and 10,000. **c**, The average stress in the notched sample under the initial energy release rates of 980 and 1030 J/m$^2$. **d**, The measured crack extension per cycle d$c$/d$N$ versus the applied energy release rate $G$ in log scale from the sample under the initial energy release rate of 1030 J/m$^2$. Under the initial energy release rate of 980 J/m$^2$, no crack extension can be observed over 10,000 cycles under a microscope with pixel resolution of 20 μm. Under the initial energy release rate of 1030 J/m$^2$, the minimum rate of crack propagation is 0.01 μm per cycle. Scale bars in **a** and **b** are 1 mm.

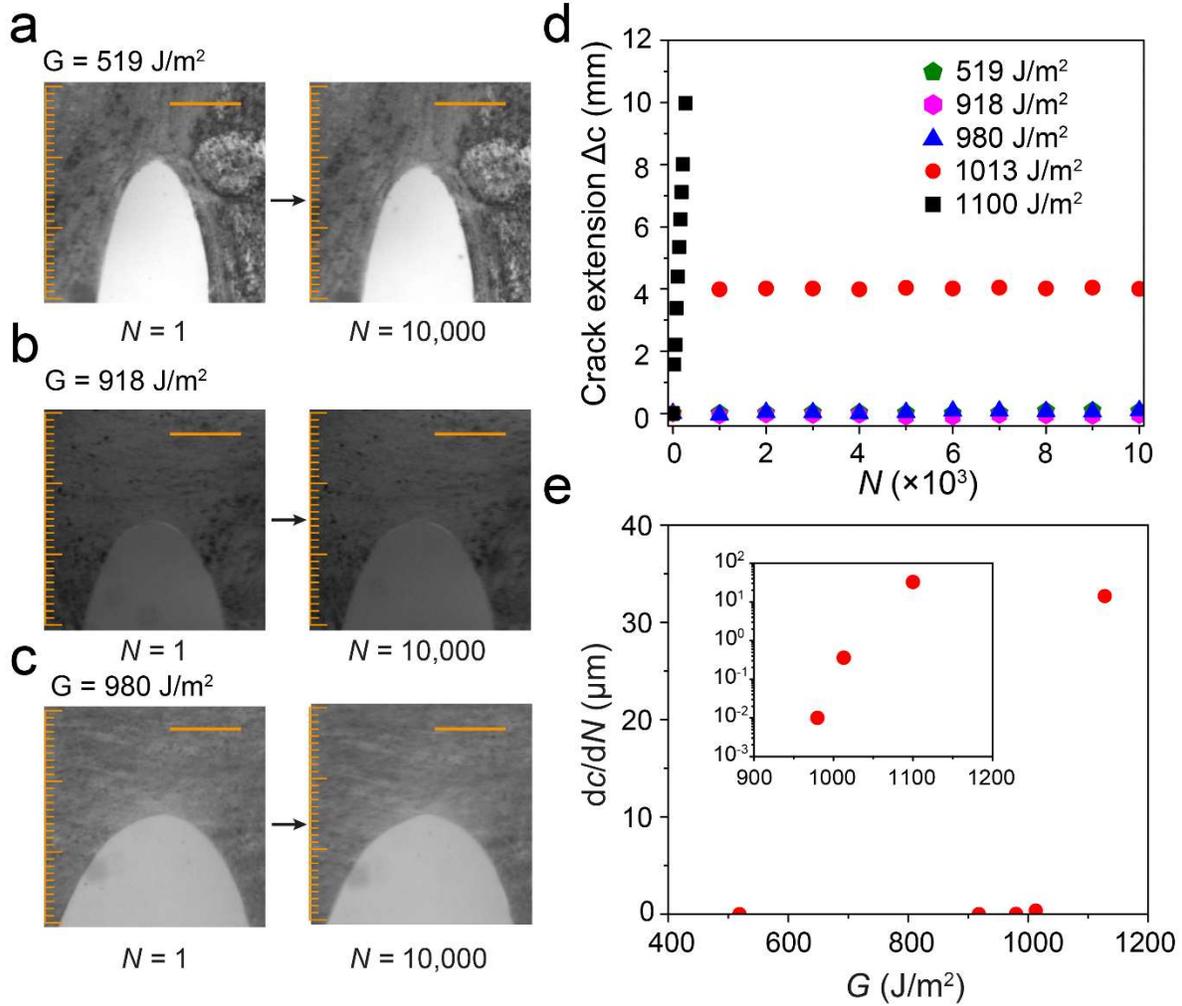

**Fig. S6. Validation of high fatigue threshold with pure shear test.** Images of the notched sample for 90-min dry-annealed PVA at the 1st cycle and 10,000th cycle under the applied energy release rates of **a**, 519 J/m$^2$, **b**, 918 J/m$^2$ and **c**, 980 J/m$^2$. **d**, Crack extension $\Delta c$ versus cycle number $N$ under the applied energy release rates of 519, 918, 980, 1013 and 1100 J/m$^2$. **e**, Crack extension per cycle d$c$/d$N$ versus the applied energy release rate $G$ in linear scale. The inset plot shows the same d$c$/d$N$ for the applied energy release rates of 980, 1013 and 1100 J/m$^2$ in log scale. Under the energy release rate of 519 and 918 J/m$^2$, no crack extension can be observed over 10,000 cycles under a microscope with pixel resolution of 20 μm. Under the energy release rate of 980, 1013 and 1100 J/m$^2$, the crack extension per cycle is 0.01, 0.36 and 32 μm per cycle, respectively. Scale bars in **a**, **b**, and **c** are 1 mm.

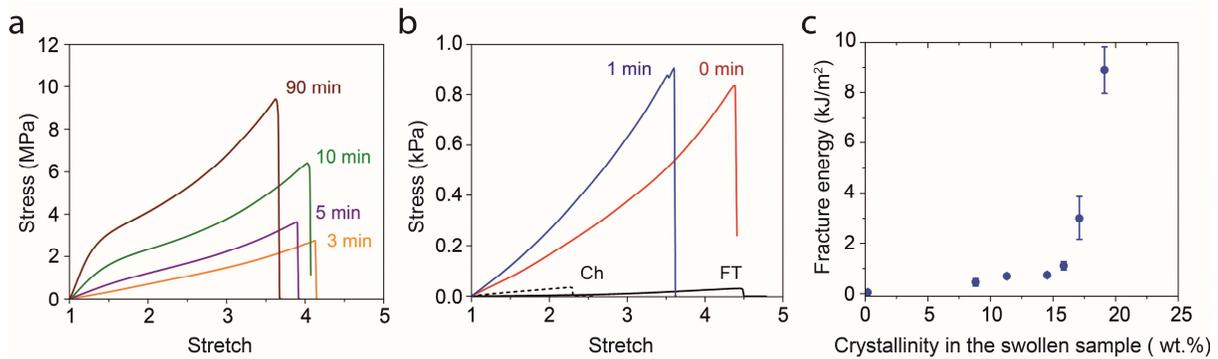

**Fig. S7**. **Mechanical characterization of PVA hydrogels with various crystallinities**. **a**, Nominal stress vs. stretch curves of dry-annealed PVA with annealing time of 3, 5, 10 and 90 min. **b**, Nominal stress vs. stretch curves of chemically crosslinked PVA, freeze-thawed PVA and dry-annealed PVA with annealing time of 0 and 1 min. **c**, Fracture energy vs. crystallinity in the swollen state.

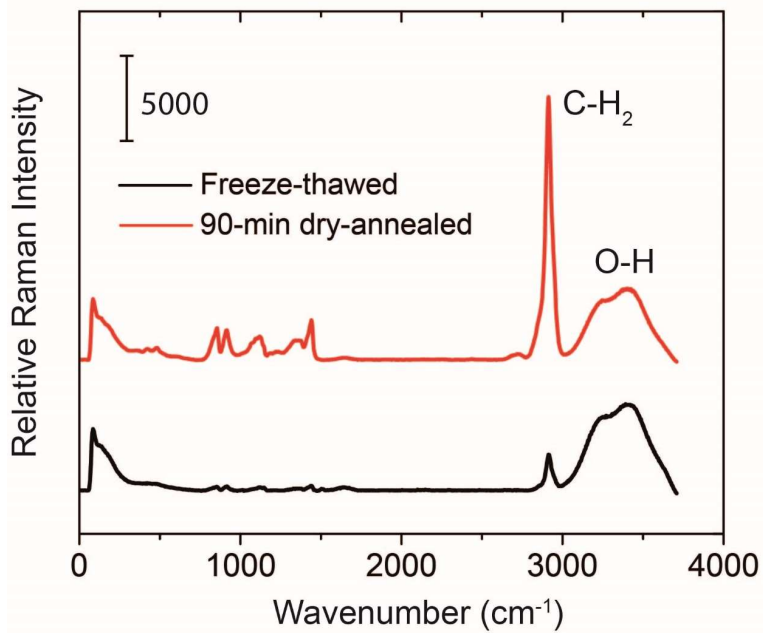

**Fig. S8**. **Raman spectroscopy of freeze-thawed PVA and 90-min dry-annealed PVA.** The pronounced intensity in the range of 3000-3500 cm$^{-1}$ originates from O-H-rich domains and the intensity in the range of 2800-3000 cm$^{-1}$ corresponds to C-H$_2$.

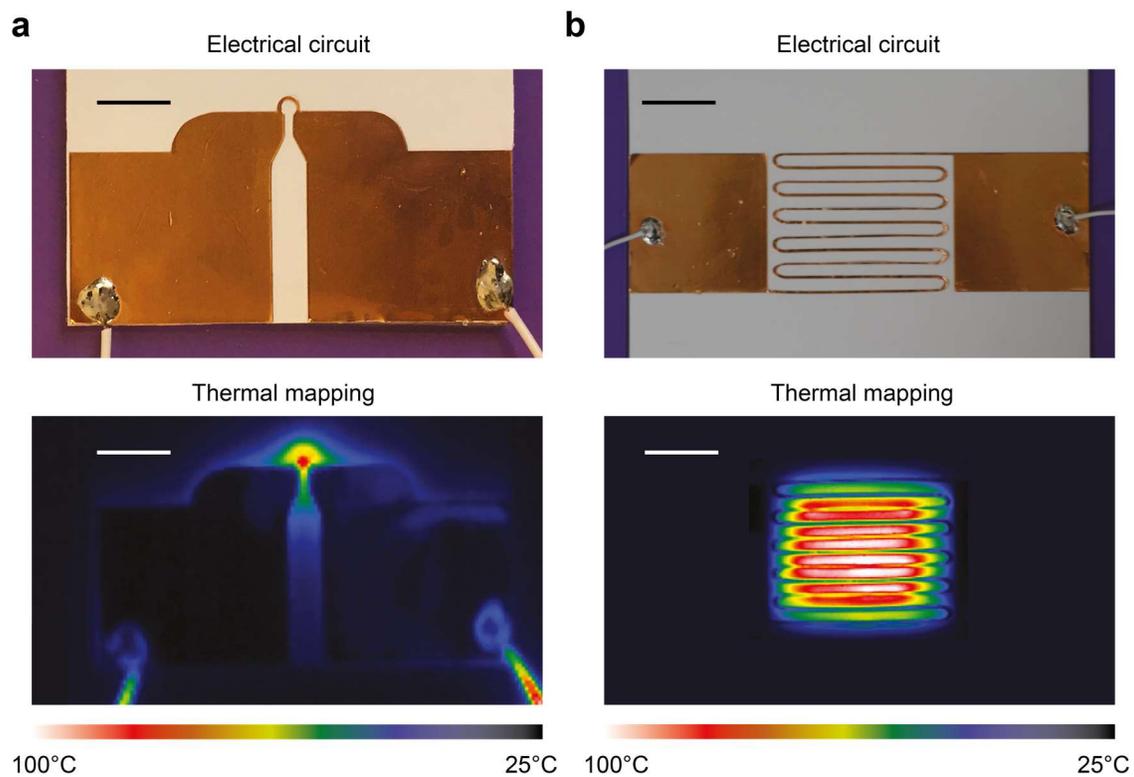

**Fig. S9**. **Electrical circuit and thermal mapping for programmable crystalline domains**. **a**, tip-reinforced sample and **b**, mesh-reinforced sample. Scale bars in **a** and **b** are 5 mm.

**Legends for videos**

**Movie S1. Tension of the pristine notched sample and the tip-reinforced sample.** The pristine sample fractures at the stretch of 1.2 with the fracture energy of 22 J/m$^2$. The tip-reinforced sample can be stretched to 1.5 without rupture and reaches the fracture energy of 300 J/m$^2$.

**Movie S2. Cyclic loading of the reinforced kirigami hydrogel sheet.** The pristine kirigami sheet is under single cycle of load with the ultimate stretch of 1.7 and ultimate effective nominal stress of 0.9 kPa. The reinforced kirigami sheet is under cyclic loading over cycles. The steady-state effective nominal stress is 26 kPa.